# The Narrow Pulse Approximation and Long Length Scale Determination in Xenon Gas Diffusion NMR Studies of Model Porous Media

R. W. Mair*[†], P. N. Sen[†‡], M. D. Hürlimann[‡], S. Patz[§], D. G. Cory[†], R. L. Walsworth*

* Harvard-Smithsonian Center for Astrophysics, Cambridge, MA 02138, USA.

[†] Dept. of Nuclear Engineering, Massachusetts Institute of Technology, Cambridge, MA 02139, USA

[‡] Schlumberger-Doll Research, Ridgefield, CT 06877, USA.

[§] Dept. of Radiology, Brigham and Women's Hospital, Boston, MA 02115, USA

**Running Title**: Narrow Pulse Effects in Xenon Gas Diffusion

**Corresponding Author:**

Ross Mair

Email: rmair@cfa.harvard.edu

# ABSTRACT

We report a systematic study of xenon gas diffusion NMR in simple model porous media: random packs of mono-sized glass beads, and focus on three specific areas peculiar to gas-phase diffusion. These topics are: (i) diffusion of spins on the order of the pore dimensions during the application of the diffusion encoding gradient pulses in a PGSE experiment (breakdown of the 'narrow pulse approximation' and imperfect background gradient cancellation), (ii) the ability to derive long-length scale structural information, and (iii) effects of finite sample size. We find that the time-dependent diffusion coefficient, $D(t)$, of the imbibed xenon gas at short diffusion times in small beads is significantly affected by the gas pressure. In particular, as expected, we find smaller deviations between measured $D(t)$ and theoretical predictions as the gas pressure is increased, resulting from reduced diffusion during the application of the gradient pulse. The deviations are then completely removed when water $D(t)$ is observed in the same samples. The use of gas also allows us to probe $D(t)$ over a wide range of length scales, and observe the long-time asymptotic limit which is proportional to the inverse tortuosity of the sample, as well as the diffusion distance where this limit takes effect (~ 1 - 1.5 bead diameters). The Pade′ approximation can be used as a reference for expected xenon $D(t)$ data between the short and long time limits, allowing us to explore deviations from the expected behaviour at intermediate times as a result of finite sample size effects. Finally, the application of the Pade′ interpolation between the long and short time asymptotic limits yields a fitted length scale (the "Pade′ length"), which is found to be ~ $0.13b$ for all bead packs, where $b$ is the bead diameter.

**Keywords**: xenon, diffusion, PGSE, tortuosity, porous media

# INTRODUCTION

NMR is a commonly used, non-invasive probe of *liquid*-saturated porous materials. The $^1$H signal obtained from water-saturated rocks and model porous systems can provide a wealth of information including the pore surface-area-to-volume ratio ($S/V_p$) (*1,2*), the average pore size in model systems (*3,4,5*), the eccentricity of non-spherical pores (*6*), and visualizations of fluid transport under flow (*7*). The restricted diffusion of water molecules is commonly studied, and the Pulsed Gradient Spin Echo (PGSE) technique has become a powerful tool for this purpose (*8,9*). However, this approach is usually limited to probing length scales < 20 µm, because spin relaxation quenches the NMR signal before molecules can diffuse across longer distances. For this reason, liquid phase NMR has been unable to yield measurements of long-range properties for most porous media; e.g., heterogeneous length scales and tortuosity in reservoir rock samples (*2,10*). Tortuosity is an important parameter that characterizes pore connectivity and fluid transport properties in porous media such as electrical conductivity, fluid flow, diffusion, and velocity of sound (*11*).

We have recently shown that *gas* diffusion NMR can be a powerful probe of porous media (*12,13*). The spin 1/2 noble gases ($^3$He and $^{129}$Xe) are particularly well suited for such studies, given their rapid diffusion, inert nature, low surface interactions which reduce surface $T_1$ effects, and the ability to tailor the diffusion coefficient by altering the gas pressure in the sample. For example, $^{129}$Xe gas has an unrestricted diffusion coefficient, $D_0 = 5.7 \times 10^{-6}$ m$^2$ s$^{-1}$ at 1 bar pressure (*12*), which is ~ 3 orders of magnitude higher than that of water. We have used $^{129}$Xe gas diffusion NMR to extend the length scales that can be probed in porous media by more than an order of magnitude. In particular, we have shown that pore length scales on the order of millimeters, rather than tens of micrometers, can be probed; and that gas diffusion NMR can be used to measure the tortuosity of random bead packs and reservoir rocks (*13*).

In this paper, we address three key issues that arise from gas-phase NMR diffusion measurements in porous media, regardless of whether the gas signal results from thermal equilibration in the applied magnetic field, or the increasingly popular laser-polarization process (*14*). The first of these issues is the effect of diffusion during the application of the diffusion-encoding gradient pulse (*15-18*). While the large gas diffusion coefficient is the essential property that permits the probing of long length scales, it also implies that gas spins can diffuse significantly across the enclosing pore space during the application of the diffusion-encoding gradient pulses. Such a phenomenon violates a common assumption in the PGSE experiment, i.e., the *narrow pulse approximation* which assumes the gradient pulse is infinitely narrow so that the spins do not diffuse significantly during its application (*9*). While the impact of a breakdown of the narrow pulse approximation has previously been studied theoretically (*15-18*), it is only with gas-phase diffusion that the phenomenon is readily observed, and at times, unavoidable. Such significant diffusion during the application of the gradient pulses also implies that the cancellation of background gradients resulting from susceptibility differences in the sample may not be effective. This cancellation is achieved by using a modified PGSE sequence that employs bi-polar gradient pulse pairs separated by a 180° RF pulse rather than a single diffusion encoding pulse, and relies on the spins being in the same background gradient during the application of the two bi-polar pulses (*19,20*). Our past experiments show deviations in normalized time-dependent gas diffusion coefficients, $D(t)/D_0$, from values expected from the known pore surface-area-to-volume ratio ($S/V_p$) of model packed bead samples (*13*), which we have attributed to a breakdown of the narrow pulse approximation. In this paper, we report systematic measurements of $D(t)/D_0$ as a function of diffusion length during the gradient pulse over a wide range of bead sizes, and show that a reduced free diffusion coefficient, $D_0$, either by increased gas pressure or changing the observation spin to water, can indeed reduce or eliminate the previously observed deviations of $D(t)/D_0$ from the expected values.

The other two issues we address are the ability to derive intermediate and long-length scale structural information from the time-dependent diffusion coefficient, $D(t)$, and the effect of finite sample size. In gas-phase $D(t)$ experiments, the tortuosity limit is easily observed, allowing not only accurate determination of the tortuosity of a sample, but the length-scale at which the pore-structure becomes homogeneous. In previous water $D(t)$ studies in small beads, it has been shown that all $D(t)/D_0$ data could be well described by a Pade′ approximation which interpolates between the short and long time limits (*1,2*). In this paper, we therefore use the Pade′ form as convenient reference to assess factors influencing the observed gas $D(t)/D_0$ data in the intermediate time range. In particular, we observed that intermediate-time $D(t)/D_0$ data deviated from the Pade′ approximation when the sample size was within an order of magnitude of the bead size, while in a larger sample cell these deviations were eliminated. Use of the Pade′ interpolation method also involves fitting an adjustable length scale parameter which we refer to as the Pade′-length, and which we found to be a constant fraction of the bead diameter for all samples.

# THEORY

The NMR echo signal observed in a PGSE experiment has a Fourier relationship to the probability of spin motion – the so-called displacement propagator, which can be thought of as a spectrum of motion. The echo signal, $E$, obtained in a PGSE experiment for a system with uniform initial polarization, and when the narrow pulse approximation is valid, can thus be written as (9):

$$E(\mathbf{q},t) = \int \rho(\mathbf{r}) \int P_s(\mathbf{r}|\mathbf{r'},t) \times \exp[i\mathbf{q}\cdot(\mathbf{r'}-\mathbf{r})] d\mathbf{r'}\, d\mathbf{r}, \qquad [1]$$

where $\rho(\mathbf{r})$ is the spin density at position $\mathbf{r}$, and $P_s(\mathbf{r}|\mathbf{r'}, t)$ is the displacement propagator, or the probability of a spin having a displacement $\mathbf{r'} - \mathbf{r}$ during the 'diffusion time' $t$ (often referred to as $\Delta$ in the literature). However, in these experiments, the propagator $P_s(\mathbf{r}|\mathbf{r'}, t)$ depends both on the initial and final positions, not just the displacement vector $(\mathbf{r'} - \mathbf{r})$, as the samples are not translationally invariant, due to their finite size. Therefore, it is not possible in this case for the propagator to represent an average over all similar displacements $\mathbf{r'} - \mathbf{r}$. $\mathbf{q}$ is the wavevector of the magnetization modulation induced on the spins by a field gradient pulse of strength $g$ and pulse duration $\delta$. The magnitude of $\mathbf{q}$ is given by $q = \gamma \delta g$, where $\gamma$ is the spin gyro-magnetic ratio in rad/s T. In the limit of small $\mathbf{q}$, the echo will be attenuated by a factor $\exp(-q^2 D(t)\, t)$ (9), where $D(t) = \langle [\mathbf{r'}-\mathbf{r}]^2 \rangle / 6t$ is the time-dependent diffusion coefficient describing incoherent random motion of the spins in the pore space. $D(t)$ can vary as a function of $t$, with more spins encountering barriers to their motion as $t$ is increased.

For short diffusion times (i.e., small $t$), the fraction of spins whose motion is restricted by pore boundaries is given by the ratio of the volume of a boundary layer to the total pore volume, i.e., $\sim (S/V_p)\sqrt{D_0 t}$, where $S/V_p$ is the pore surface-area-to-volume ratio and $\sqrt{D_0 t}$ is the characteristic free-spin-diffusion length for a diffusion time $t$. Using this basic physical picture, Sen and co-workers have calculated, for small $t$ (21,22):

$$\frac{D(t)}{D_0} = 1 - \frac{4}{9\sqrt{\pi}} \frac{S}{V_p} \sqrt{D_0 t} + O(D_0 t). \qquad [2]$$

This relation has been verified experimentally (1) with NMR of liquids imbibed in model porous media consisting of random packs of mono-sized spherical beads, where $S/V_p = 6(1-\phi)/(\phi b)$, $\phi$ is the sample's porosity (the fraction of void space in the total volume of the sample), and $b$ is the bead diameter. We have shown this relation also holds for gas diffusion in samples with large pores (> 1 mm) (12,13). In this paper, we address short time deviations from the above description, due to a breakdown of the narrow pulse approximation for gas diffusion NMR on samples with small pores.

In addition to short-time effects, we report in this paper studies of long length-scale properties of porous media that are only observable with NMR by using gas diffusion. In particular, the tortuosity, $\alpha$, is a fundamental geometrical parameter that controls macroscopic fluid transport in porous media (23). In simple, idealized systems, $\alpha$ can be calculated: e.g., in dense random packs of spherical beads, $1/\alpha \sim \sqrt{\phi} \sim 0.62$ (24). However, in most systems such as reservoir rocks and lung tissue, $\alpha$ must be determined by measurement. In cases where surface electrical conductivity is negligible, the tortuosity of a porous medium can be determined by conductivity measurements of salt-water infused in the pore space; gas diffusion NMR is the only other static method for measuring tortuosity for systems with pore sizes > 20 μm (13). In addition, gas diffusion NMR permits one to determine how far the spins must diffuse in order to reach the tortuosity limit, which corresponds to the largest structural length scale in the pore geometry ($L_{macro}$). No other method for tortuosity determination provides this additional information.

Ignoring the effects of spin relaxation at pore boundaries (a good approximation for noble gas NMR in many systems), $D(t)/D_0$ asymptotically approaches $1/\alpha$ in the long time limit according to (25):

$$\frac{D(t)}{D_0} = \frac{1}{\alpha} + \frac{\beta L_{macro}^2}{D_0 t} - O\left(t^{-3/2}\right), \quad [3]$$

where $\beta$ is a length independent geometrical factor; implying for $\sqrt{D_0 t} \gg L_{macro}$, $D(t)/D_0 \rightarrow 1/\alpha$. Some water diffusion NMR measurements in reservoir rocks have been consistent with $D(t) \sim 1/t$ in the intermediate time regime (i.e., before the tortuosity limit is reached) (2). However, such measurements have not provided an accurate determination of $\alpha$ and $L_{macro}$. As a result, in analyzing diffusion in reservoir rocks and model porous media, it has been common to use the Pade′ approximation to interpolate between the short time ($S/V_p$) limit (Eq. [2]) and the long time ($1/\alpha$) asymptotic limit provided by electrical conductivity or calculation (2). This method uses an adjustable fitting parameter, $\theta$, which has units of time and is expected to scale with the square of the pore size (1). The Pade′ approximation can be used to estimate $D(t)$ in the intermediate $t$ regime, according to (1):

$$\frac{D(t)}{D_0} = 1 - (1 - \frac{1}{\alpha}) \times \frac{(4\sqrt{D_0 t}/9\sqrt{\pi})(S/V_p) + (1 - 1/\alpha)(D_0 t/D_0 \theta)}{(1 - 1/\alpha) + (4\sqrt{D_0 t}/9\sqrt{\pi})(S/V_p) + (1 - 1/\alpha)(D_0 t/D_0 \theta)}. \quad [4]$$

This method has provided an excellent fit to water $D(t)/D_0$ data in the intermediate-$t$ regime in model porous media consisting of small glass bead packs, with the fit parameter (the Pade′ length) yielding $\sqrt{D_0 \theta} \sim 0.14 b$, where $b$ is the bead diameter (1). In this paper, we use gas diffusion NMR to test the accuracy of the Pade′ approximation (Eq. [4]) for $D(t)$ data ranging from the short to the long time limits and with both $S/V_p$ and $\alpha$ determined by NMR.

## EXPERIMENTAL

We prepared samples of randomly packed spherical glass beads for thermally polarized xenon NMR experiments. Cylindrical glass chambers of volume ~ 50 cm³ held the bead-packs. Each sample contained beads of a single size, with bead diameters in different samples ranging from 0.1 mm to 4 mm. Also, one larger chamber cell of ~ 150 cm³ was used for additional experiments with 4 mm beads to study finite sample cell effects. In addition to the chamber for beads, all these sample cells had a second chamber of volume ~ 25 cm³, which was connected to the larger bead chamber via glass tubing obstructed by a 50 µm porosity sintered glass frit. This design allowed the gases to fill both chambers with a uniform partial pressure for the gas mixture, while the beads were confined to one chamber. We filled the samples with gas by freezing appropriate amounts of xenon (isotopically enriched to 90% $^{129}$Xe) and oxygen inside the cells at liquid nitrogen temperature. We then sealed each glass sample cell and warmed it to room temperature, establishing the desired gas partial pressures, typically ~ 6 - 6.5 bar xenon and ~ 1.5 bar oxygen. Paramagnetic oxygen was included with the xenon in order to reduce the $^{129}$Xe $T_1$ from tens of seconds to ~ 1.5 s, thereby enabling efficient signal averaging which is essential due to the low NMR signal expected with thermally polarized $^{129}$Xe gas.

The two-chamber design of the sample cells allowed the smaller, free-gas chamber to be inserted in the RF coil separately from the bead-filled chamber. Simple characterization experiments on the free gas chamber yielded the $^{129}$Xe $T_1$ time and the xenon free gas diffusion coefficient, $D_0$, for the gas mixture. We employed standard inversion-recovery (26) and PGSE (8) methods, respectively, for these measurements. A linear regression of measurements of $^{129}$Xe $T_1$ in the presence of $O_2$ in various carefully prepared standard free-gas samples of $^{129}$Xe at pressures of ~ 1 - 4 bar was performed previously (see Fig. 1). We obtained a simple empirical relation between the $^{129}$Xe $T_1$ value and the $O_2$ content ($pO_2$) in the gas mixture, valid for $^{129}$Xe pressures of a few bar near room temperature in an applied field of 4.7 T. We found $(^{129}Xe\ T_1)^{-1} = 0.0333 + 0.348 \times pO_2$, with $T_1$ in seconds and $pO_2$ in bar, and an $R^2 = 0.999$. This result compares well with the proportionality constant of 0.388 for $^{129}$Xe $T_1$ at 4.7 T as a function of oxygen density at STP, as determined previously by Jameson et. al. in their detailed analysis of intermolecular dipolar coupling in the gas phase (27). We used our relationship to determine the $pO_2$ in each filled bead cell.  We then determined $p$Xe, the xenon gas partial pressure, from simple addition of gas partial pressures and their diffusion coefficients (28):

$$\frac{1}{D_0} = \frac{p\text{Xe}}{D_0(\text{Xe})} + \frac{pO_2}{D_0(O_2)} \qquad [5]$$

where $D_0(\text{Xe}) = 5.71 \times 10^{-6}$ m$^2$ s$^{-1}$ (12) is the Xe diffusion coefficient at 1 bar pressure and room temperature and $D_0(O_2) = 13.5 \times 10^{-6}$ m$^2$ s$^{-1}$ (28) is the xenon diffusion coefficient at infinite dilution in $O_2$ at 1 bar pressure and room temperature.  We measured $D_0$ from each sample cell's free gas chamber and used it to normalize the Xe $D(t)$ measurements from the bead chamber of the corresponding cell.

Water $D(t)$ data was obtained from single chamber cells of 50 cm$^3$, similar to those used for most of the Xe $D(t)$ measurements.  We filled the cells with either 4 or 0.5 mm beads, and then saturated them with de-ionized water under vacuum conditions to ensure the removal of air bubbles.  The 1 mm beads were held in a 4 cm ID plastic cylinder, which was also water saturated in the same manner.  In all cases the beads were the same as those used in the Xe gas experiments.  We measured the water $T_1$ time using the inversion recovery method, and found it to be ~ 1.5 - 2 s without needing to add a doping agent.  No free diffusion coefficient ($D_0$) measurement was made on these cells.  Instead, we determined the water $D_0$ by extrapolation to $t = 0$ of the very linear plot of $D(t)$ vs $\sqrt{D_0 t}$.

In practice, the very short spin coherence time ($T_2$) of fluid (water or gas) spins in a porous sample, and the high background magnetic field gradients that at high applied fields result from susceptibility contrast between the solid grains and the imbibed fluid (29) make the standard pulsed gradient spin echo (PGSE) technique unsuitable for measuring $D(t)$.  Instead, we used a modified stimulated echo sequence (PGSTE-bp) incorporating alternating bi-polar diffusion-encoding gradient pulses which cancel out the effect of the background gradients on the diffusion measurement (19,20).  The sequence is illustrated in Fig. 2, where the timing parameters shown correspond to the description in the previous section or are described in the figure caption.  The additional 180° RF pulses around each bi-polar gradient pulse pair provide refocusing to overcome fast inhomogeneous signal loss; while the crusher gradients, applied on an orthogonal axis, remove spurious signal resulting from imperfect 180° RF pulse calibration and $B_1$ inhomogeneity.  The gradient pulses are half-sine shaped in order to minimize their temporal length ($\delta$) while maintaining a reproducible gradient shape, and to reduce eddy current ring-down after the application of the gradient pulses.  We determined $D(t)$ by fitting the natural log of the measured echo attenuation to a modified form of the Stejskal-Tanner equation (20):

$$\ln(E(g,t)/E(0,t)) = -g^2\gamma^2\delta^2(2/\pi)^2 D(t)(t - \delta/8 - T/6) \qquad [6]$$

where we have re-expressed the echo signal, $E$, in terms of the gradient pulse strength, $g$, and time, $\delta$, and the $(2/\pi)^2$ factor accounts for the half-sine shape of the gradient pulses. $T$ and $t$ are defined in Fig. 2. This equation is only applicable in the limit of $q = \gamma\delta g > 1/L$, where $L$ is the sample length (2). As a result, we omitted data from very low strength gradient pulses from all fits. However, we ensured the gradient pulse strengths were not so high as to leave the Gaussian, low-$q$ limit for which measured $\ln(E(g,t)/E(0,t))$ values are proportional to $g^2$. Deviations from this Gaussian limit usually occurred around a maximum signal attenuation of ~ 20%, as has been previously noted (2).

We implemented the PGSTE-bp technique on two commercial NMR spectrometers. For most Xe experiments, we used a GE Omega/CSI spectrometer (GE NMR Instruments, Fremont, CA), interfaced to a 4.7 T magnet with clear bore of 20 cm. Experiments were performed at 55.35 MHz for $^{129}$Xe using a tuned home-made solenoid RF coil. Applied magnetic field gradients up to 7 G/cm in strength were available on this system. For xenon experiments on the 4 mm beads in the larger chamber cell, and all water experiments, we used a Bruker AMX2 - based spectrometer (Bruker Instruments Inc., Billerica, MA) interfaced to a 4.7 T magnet. This system is equipped with a 12 cm ID gradient insert (Bruker) capable of delivering gradient pulses of up to 26 G/cm. We employed Alderman-Grant-style RF coils, tuned to 200.4 MHz for $^{1}$H and 55.35 MHz for $^{129}$Xe (both supplied by Nova Medical Inc., Wakefield, MA). All experiments were performed at room temperature.

For the PGSTE-bp measurements of Xe $D(t)$ on both instruments, we set $\delta$ to 750 µs, the minimum possible time without distortion of the gradient shape on the GE system. We varied the diffusion time, $t$, from a minimum of 25 ms to a maximum of 3 s. We chose gradient strengths, $g$, to produce significant attenuation of the $^{129}$Xe NMR signal while remaining between the high and low $q$ regimes described earlier. The combined timing parameters resulted in a diffusion encoding time $T \sim 4$ ms in all experiments. We employed 12 different $g$ values and a repetition time of ~ 7-8 s, while the number of signal averaging scans ranged from 16 - 256, depending on signal linewidth, signal strength, and diffusion time $t$. For water $D(t)$ measurements, we set $\delta$ to 1 ms, and varied the diffusion time, $t$, from 30 ms to 4 s. We chose 16 gradient strengths, a repetition rate of ~ 8 - 12 s, and acquired 8 - 32 signal averaging scans. The diffusion encoding time $T \sim 8 - 18$ ms.

# RESULTS

Fig. 3 provides an overview of the Xe $D(t)$ data at ~ 6 bar Xe pressure, from six different samples each with a different bead size. The data is plotted in terms of the reduced diffusion coefficient, $D(t)/D_0$, as a function of normalized diffusion length, $b^{-1}\sqrt{D_0 t}$, where $D_0$ is the free gas diffusion coefficient and $b$ is the bead diameter. Using this method of display, it is expected that the $D(t)/D_0$ data from all random bead packs should collapse onto a single "universal" trend (*1*). The data is shown on two separate plots (Figs. 3a and 3b) due to the wide range of bead sizes, and hence normalized diffusion lengths, which vary by a factor of 40. Fig. 3a covers the larger bead sizes, 1 to 4 mm, where smaller normalized diffusion distances are possible, and the short-time $S/V_p$ regime can be probed carefully. Fig. 3b shows the data for the smaller beads, 0.1 to 1 mm, where it is possible to monitor Xe diffusion over many bead diameters, and the asymptotic value of $D(t)$ at the tortuosity limit is easily observed. The expected universal $S/V_p$ and tortuosity limits for $D(t)$ in random bead packs are shown on the figures in dashed and solid straight lines respectively. The large bead data agree well with the short-time $S/V_p$ limit; whereas for smaller beads, the short-time data deviates from the calculated $S/V_p$ line, as seen previously (*13*). Conversely, the measurements for all bead sizes agree with the calculated long-time limit to ~ 5%, where deviations likely arise from variations in packing, and hence porosity, from sample to sample. Deviations are particularly evident in the small bead samples in Fig. 3b, where the observed inverse tortuosity $1/\alpha$ ~ 0.66 corresponds to ~ 10% greater porosity, which may arise from non-optimum packing or static-charge-enabled clumping.

Fig. 4 shows the Xe $D(t)$ measurements in more detail. The 6 bar Xe pressure $D(t)/D_0$ data from each sample is plotted separately in Fig. 4 a-f, along with the corresponding $D(t)/D_0$ data obtained from similar samples filled with 3 bar Xe - these data points are reproduced from ref. (*30*). Note that the 6 bar Xe data has better SNR than the previously acquired 3 bar Xe data because of the higher Xe pressures, the use of isotopically enriched $^{129}$Xe, and longer signal averaging. Also included in the plots for 0.1, 0.5, 1 and 4 mm beads is water $D(t)/D_0$ data we acquired (0.5, 1, 4 mm) or reproduced from ref (*1*) (0.1 mm). These plots show the potential for combined water and gas phase $D(t)$ measurements to provide complementary information over a range of length scales, from the short-$t$ $S/V_p$ regime, for which the gas phase $D(t)$ method fails with very small pores, to the long-$t$ tortuosity regime, which cannot generally be probed with the water $D(t)$ measurements.

Each plot in Fig. 4 shows the observed tortuosity limit derived for the 6 bar Xe data, and is seen to differ for each sample. For samples with 3 mm beads or smaller, where the $1/\alpha$ limit is clearly observed, we used the knowledge that $1/\alpha \sim \sqrt{\phi}$ for random bead packs, where $\phi$ is the porosity, to calculate a unique short-$t$ $S/V_p$ line for each sample using $S/V_p = 6(1-\phi)/(\phi b)$ (*1*). This provided an indication of the short-time $S/V_p$ limit in the small beads, where it is not necessarily observed. However, for the 4 mm beads, the observed $D(t)/D_0$ at both pressures lies exactly on the expected $S/V_p$ line at short diffusion times, while it is not clear that the correct tortuosity limit has been reached since the data asymptotes at significantly shorter diffusion lengths than for smaller bead samples. Hence for the 4 mm beads we estimated the inverse tortuosity, $1/\alpha$, from the observed $S/V_p$ line and the value of $\phi$ this yielded, rather than using the observed $1/\alpha$ to estimate $S/V_p$ as was done for all other data sets. In addition, we used the Pade′ approximation, Eq. [4], to interpolate between the long and short $t$ limits for each sample (*1*). Because one goal of this study is to compare the Pade′ approximation to intermediate-$t$ $D(t)/D_0$ measurements, we adjusted the Pade′ length, $\sqrt{D_0 \theta}$, for each sample to provide the best fit with the short-$t$ and long-$t$ limits, i.e., the dashed and solid lines in each plot in Fig. 4. In other words, we interpolated the limiting behavior at short and long times using Eq. [4], rather than attempting to fit the intermediate-$t$ experimental data itself. The Pade′ approximation for interpolating between the short-$t$ and long-$t$ limits provides a theoretical basis for determining unknown (or using known) structural parameters. We do not attempt to fit lines through all our observed data in all circumstances if the fitting procedure does not yield useful structural information. Table 1 summarizes the information about bead pack structure which we determined from the 6 bar Xe data, along with that obtained for smaller beads from previous water $D(t)$ NMR measurements (*1*).

Finally, Fig. 5 shows additional Xe $D(t)/D_0$ data from a 4 mm bead sample acquired in a larger (~ 150 cm$^3$) sample holder. The data is contrasted to that obtained from the ~ 50 cm$^3$ sample cell (Fig. 4a), showing the effect of finite sample size on $D(t)/D_0$: in smaller samples the tortuosity limit is reached earlier, $L_{macro}$ is systematically underestimated, and deviations from the Pade′ line are observed.

## DISCUSSION

The experiments reported in this paper address three key issues in gas diffusion NMR: (i) atomic motion during the application of diffusion-encoding gradient pulses, particularly breakdown of the narrow pulse approximation; (ii) the determination of intermediate and long length scale structural information; and (iii) finite sample size effects. Here, we summarize conclusions about these issues that can be drawn from our studies of xenon gas diffusion NMR in random packs of mono-sized beads.

The 1D Xe diffusion distance during the application of the gradient pulse is $= \sqrt{2D_0\delta}$. Taking the measured free gas diffusion coefficient, $D_0 = 1.36 \times 10^{-6}$ m$^2$ s$^{-1}$ (*12*), for the xenon-oxygen mixture used for the 3 bar Xe experiments, the 1D diffusion distance ~ 45 μm during the spectrometer's minimum gradient pulse time of $\delta = 750$ μs. Additionally, the characteristic length scale of the pore space between packed beads is approximately 1/4 the bead diameter. Therefore, during the diffusion encoding gradient pulse, xenon atoms at 3 bar pressure traverse ~ 36% of the pore in a 0.5 mm bead pack and ~ 18% of a pore in a 1 mm bead pack. Fukushima and co-workers have suggested (*31*) that the narrow pulse approximation fails when the diffusion distance during the gradient pulse is greater than about 14% of the pore size. Thus it is no surprise that we encountered a failure of the narrow pulse approximation which is manifested as the $D(t)/D_0$ data deviating from theoretical predictions for 3 bar xenon pressure in samples with beads of 1 mm diameter or smaller (see Fig. 4d-f). The narrower necks that run off the pores between the beads may also induce similar but less dramatic effects in larger beads (i.e., 3 bar xenon in 2 mm beads, Fig. 4c)

Increasing the xenon pressure to ~ 6.5 bar reduced $D_0$ to ~ $0.80 \times 10^{-6}$ m$^2$ s$^{-1}$ and thereby reduced the diffusion distance during the gradient pulse. We observed a clear effect of the higher gas pressure in the $D(t)/D_0$ values measured at short times for both the 1 and 0.5 mm beads (Fig. 4d-e): the $D(t)/D_0$ data moves down and closer to the Pade′ and $S/V_p$ lines, indicating a less severe breakdown of the narrow pulse approximation. Hence it is clear that diffusion during the gradient pulse on a length scale approaching that of the characteristic pore length is a significant contributor to deviations in short-time $D(t)/D_0$ measurements. Note also that water diffusion NMR provides an accurate measure of $S/V_p$. This technique is thus complimentary to gas diffusion NMR for systems with pore sizes < 200 μm, for which elimination of significant diffusion during the gradient pulse through increased gas pressure is impractical (see Fig. 4d - 4f). The short-time discrepancy between the water and xenon $D(t)/D_0$ data seen in Figs. 4d and 4e are consistent with previous water and gas-phase $D(t)/D_0$ measurements in reservoir rocks (*13*).

Another important systematic effect of significant atomic motion during the diffusion-encoding gradient pulses is a reduced effectiveness of the bi-polar stimulated echo (PGSTE-bp) sequence to cancel background gradients. Nominally, this sequence cancels susceptibility-induced background gradients by the application of two gradient pulses of opposite sign, separated by a 180° RF pulse, which results in cumulative addition of the dephasing effect of the gradient pulses and subtraction of the (assumed constant) background gradient (*19,20*). This cancellation relies on each spin experiencing the same background gradient during the application of both pulses in a bi-polar pair. In the small beads, $\sqrt{D_0\delta}$ → pore size, hence there is a high probability that the spins experience different background gradients during the two gradient pulses. In addition, for beads of different sizes but made with the same materials so they have identical magnetic properties, the susceptibility-induced background gradients are inversely proportional to the bead size. Therefore, as the background gradient at any place in the pore is a function of the distance from the pore surface, in the small beads there is a greater chance that the variation in background gradient experienced by a spin during the application of the bi-polar pulses will be significant. These two effects - significant diffusion across the pore space and larger background gradients - reduce the effectiveness of the PGSTE-bp sequence in canceling background gradients in small bead samples. This reduced effective cancellation of background gradients causes the xenon spins to experience a larger effective applied gradient, which results in greater signal attenuation and hence an artificially large observed $D(t)$, as is seen in small beads in the short time limit. In general then, as $D_0$ (and thus $\sqrt{D_0\delta}$) increases, the deviation in $D(t)/D_0$ for a given diffusion distance increases, with such deviation being greater for smaller beads.

The fact that increasing $\sqrt{D_0\delta}$ results in increased values of $D(t)/D_0$ for a given diffusion distance is an interesting observation, and is the opposite of what can be concluded from theoretical predictions for $E(q)$ vs $q$ as a function of $\sqrt{D_0\delta}$ in the case of restricted diffusion (*15-17*). It has been shown by simulation that for PGSE experiments on spins in restricted environments, the observed echo attenuation minima (and 'diffusive diffraction' peaks (*3*) ) on a plot of $\ln(E(q)/E(0))$ vs $q$ move to higher values of $q$ as $\sqrt{D_0\delta}$ → the pore size, thus giving the potential for misleading pore or cell sizes to be read from the $1/q$ value of the minimum. $D(t)$ is determined from the initial slope of $\ln(E(q)/E(0))$ vs $q^2$ before the signal decays one order of magnitude. In the simulations, this slope, and thus $D(t)$, *decrease* as $\sqrt{D_0\delta}$ increases. By contrast, our data in Fig. 4 shows that $D(t)/D_0$ *increases* as $\sqrt{D_0\delta}$ is increased. This discrepancy arises, we believe because these theoretical studies consider a different problem - i.e. they only consider diffusion of the spins during $\delta$, but not that the background gradients experienced by the spins may be changing at this time.

With the aim of making these different but similar, and related scenarios clearer, we have outlined three phenomena (potential systematic effects) in a flow-chart in Fig. 6, and note the consequences for $D(t)/D_0$ measurements as a result of each. The three phenomena are: 1) restricted (non-Gaussian) diffusion; 2) motion of Xe spins during $\delta$; and 3) spins experiencing background gradients which vary during $\delta$. It is well known that phenomenon 2 alone has no effect on diffusion coefficient determination if the diffusion is Gaussian (*9*). The prior theory discussed above (*15-17*) deals with the combined effect of phenomena 1 and 2, while our explanation above outlines the case for the results seen for phenomena 1 and 3. However, we believe the present experiment is one of the studies of the effects of all three phenomena in a system in which all three may be significant. We conclude that the dominant effect on the observed $D(t)/D_0$ in our restricted gas-phase system is the varying background gradients during $\delta$. Thus, $D(t)/D_0$ increases as $\sqrt{D_0\delta}$ is increased because the effect of the background gradients varying during $\delta$ is felt by all spins within a characteristic length of the pore wall. The restricted diffusion effects, which would result in $D(t)$ decreasing as $\sqrt{D_0\delta}$ increases, according to (*15-17*), is minor in comparison, as such effects are only realized when spins actually collide with the pore wall, an event of much lower probability than moving near the wall.

Clearly, the power of noble gas diffusion NMR is the ability to probe very long distance sample heterogeneity with insignificant spin depolarization, thereby permitting an accurate measurement of tortuosity, an important parameter which cannot be measured with diffusion NMR of liquid-saturated porous media with pores > 20 μm (*2,10*). It should be noted that for random bead packs, the tortuosity ($\alpha$) should be independent of bead size; however, Fig. 4 shows fine variations in tortuosity from sample to sample. We believe these variations in $\alpha$ shown in Fig. 4 result from irregularities in the dense random packing structure of the different bead packs, which were observed visually to become more random and less dense as the bead size was reduced. In addition, as expected the observed tortuosity does not change for a particular sample of beads as a function of the pressure of xenon gas used, except for the 0.1 mm beads. The variation in $\alpha$ with gas pressure in the 0.1 mm beads may be due to imperfect $D_0$ calibration in the 3 bar xenon $D(t)$ measurements, where $D_0$ was not measured independently (*30*).

It was recently suggested (*32*) that the observed long-time limit for $D(t)/D_0$ (i.e. $1/\alpha$) in a given sample geometry would be reduced with increased background gradients and imperfect cancellation of these gradients by the bi-polar gradient pulses in the PGSTE-bp sequence. Our data, however, shows incremental increases in $1/\alpha$ as the bead size gets smaller (and hence the background gradients as well as significant diffusion during the gradient pulse increase). In addition, the contention in (*32*) that increased imperfect cancellation of background gradients reduces the observed long-*t* $D(t)/D_0$ limit implies that a lower $1/\alpha$ should be observed for each sample with ~ 3 bar Xe pressure as compared to ~ 6 bar Xe pressure. However, the plots in Fig. 4 show that the determination of tortuosity is not affected by these changes in gas pressure (except for the 0.1 mm beads, for which a *greater* $1/\alpha$ is found at lower pressure). This independence of the determination of $1/\alpha$ on moderate changes in gas pressure implies that low pressure gases, or rapidly diffusing gases such as laser-polarized $^3$He, may be suitable for tortuosity measurements at long diffusion times in a variety of samples. Such a capability would be important in more complex porous media than bead packs (e.g., reservoir rocks and lung tissue) which can have very long length-scale heterogeneity.

At intermediate diffusion times, except in the case of the 4 mm beads, the $D(t)/D_0$ data generally lie close to the expected Pade′ approximation line. Despite the variations in packing, which effect the porosity, tortuosity and $S/V_p$ from sample to sample, the normalized Pade′ length, $b^{-1}\sqrt{D_0\theta}$, is quite similar for the different bead packs (see Table I), with an average value of 0.132 ± 0.005. This result is consistent with theory (*1*). (Note that the value of $b^{-1}\sqrt{D_0\theta}$ ~ 0.145 given in (*1*) for water-infused glass beads was determined using an estimated and uncertain value for the tortuosity.)

Finally, a subtle effect is also notable in gas diffusion NMR of the larger beads: the intermediate-*t* $D(t)/D_0$ data drops below the calculated Pade′ line and reaches the tortuosity limit earlier than predicted by the interpolation routine (see Fig. 4a). We believe this deviation from the Pade′ approximation is due to the finite size of the sample relative to the bead size, which can invalidate the assumption of an infinite sample used to derive Eq. [2-4] above. This finite-size hypothesis is supported by the agreement of the $D(t)/D_0$ data with the Pade approximation in the intermediate-*t* regime for a 4 mm bead pack in a large sample container (see Fig. 5). Due to the larger volume of this additional cell and variations in the geometrical configurations between the Bruker and GE spectrometers, the sample length, *L*, along the direction of diffusion encoding was ~ 3 times longer for experiments on the large cell than for the small cell. In the small cell, $L \sim 8b$ while $L \sim 25b$ in the large cell, evidently the latter condition satisfies the condition $L \gg b$.[Footnote 1]

# CONCLUSION

Gas diffusion NMR is a powerful probe of porous media. In the current work, we have used thermally polarized xenon imbibed in random packs of mono-sized glass beads to study three key issues of this technique: breakdown of the narrow pulse approximation, the determination of long length scale structural information, and finite sample size effects.

The large gas diffusion coefficient, which permits the probing of long length scales, also causes rapid diffusion during the application of the diffusion-encoding gradient pulses. When this diffusion distance approaches the pore size, the narrow-pulse approximation is violated and background gradient cancellation by the PGSTE-bp technique may not be optimal. For samples containing an NMR observable gas and pore sizes of a few hundred microns, such violations can easily occur, leading to systematic errors in the determination of the pore surface-area-to-volume ratio, $S/V_p$. We have shown that such systematic errors can be ameliorated by increasing xenon gas pressure, which in turn decreases the unrestricted diffusion coefficient, $D_0$, and hence reduces the diffusion distance $\sqrt{D_0 \delta}$ during a gradient pulse of time $\delta$. The deviations are removed entirely when water is used as the NMR-detectable medium, due to the much slower diffusion of water molecules. Breakdown of the narrow pulse approximation in gas diffusion NMR will be particularly significant when probing structures in samples where limited gas pressures are required: e.g., 1 bar pressure in lung tissue. Conversely, we found that at long diffusion times, moderate changes in gas pressure, and hence $\sqrt{D_0 \delta}$, had no effect on the observed value of $D(t)/D_0$ in most samples. The diffusion distance at which the long-distance (i.e., tortuosity) asymptote was reached was generally observed to be $\sqrt{D_0 t} \sim$ 1 - 1.5 bead diameters for all samples.

We also studied the approach of $D(t)/D_0$ to the tortuosity asymptote. We used the Pade′ approximation to interpolate between the short and long diffusion time limits, and found reasonable agreement with experimental data in the intermediate $t$ regime. The Pade′ fitting method also yielded a single fitted parameter, $\theta$, which for mono-sized bead packs is simply related to the bead diameter. Across a range of bead sizes of more than an order of magnitude, we found that the Pade′ length, $\sqrt{D_0 \theta} \sim 0.132b$, where $b$ is the bead diameter. This value is similar to but slightly lower than the Pade′ length of $0.145b$ previously determined using water $D(t)/D_0$ measurements in very small bead packs. A challenge in future work will be to relate $\sqrt{D_0 \theta}$ to structural length scales in heterogeneous systems. Finally, we found that the measured xenon $D(t)/D_0$ reached the long-time, tortuosity limit at smaller normalized diffusion distances for samples with larger beads, when the sample cell size was kept constant. We confirmed that this effect resulted from the finite sample size when the experiments were carried out in a larger sample cell.

## ACKNOWLEDGEMENTS

This work was supported by the NSF, NASA, and the Smithsonian Institution.

## FOOTNOTE

We note also that for the standard sized (smaller) sample container used to acquire all the data in Fig. 4, the deviation of $D(t)/D_0$ below the Pade′ line at intermediate times decreases and disappears as the bead size reduces below 4 mm. However, there is still evidence of a small deviation in the 2 mm bead sample at intermediate times that is not observed in the 3 mm beads. It is likely that such minor deviations result from physical properties that vary from sample to sample. This may include the glass composition and inherent paramagnetic impurity level for each type of bead - factors that will add to variability in background gradients experienced during the measurements. In addition, variations in packing geometry due to non-uniformity in grain size, shape defects and tilt can induce effects similar to those manifested by finite-sample size effects.

# FIGURE CAPTIONS

Table 1. Summary of porosity, $\phi$, inverse tortuosity, $1/\alpha$, pore surface area-to-volume ratio, $S/V_p$, Pade' fitting parameter, $\theta$, free diffusion coefficient, $D_0$, and normalized Pade' length, $b^{-1}\sqrt{D_0\theta}$ derived from Xe gas $D(t)$ data at ~ 6 bar pressure in all samples. Similar data for water $D(t)$ experiments in small beads is reproduced from Ref. (*1*).

Fig. 1. Measured $^{129}$Xe $T_1^{-1}$ as a function of $O_2$ partial pressure for samples of a few bar of Xe and $O_2$. The fitted line is a regression analysis omitting the one outlying point, and was used to derive the empirical relation between $^{129}$Xe $T_1$ and $pO_2$. Error bars are the same size or smaller than the data symbols.

Fig. 2. Pulse sequence diagram for the Pulsed Gradient Stimulated Echo (PGSTE-bp) with alternating bi-polar gradient pulses, as used in this work. The diffusion encoding gradient pulses, of length $\delta$ and strength $g$, are shown in gray, while crusher gradients are shown in black. The diffusion time is denoted $t$ and the total diffusion encoding time is $T$. See text for further description.

Fig. 3. Time-dependent diffusion measurements for thermally polarized xenon gas at ~ 6 bar pressure imbibed in randomly packed spherical glass bead samples. Each sample contains beads of a uniform diameter. The $^{129}$Xe time-dependent diffusion coefficient is normalized to the free gas diffusion coefficient, $D_0$, and plotted versus normalized diffusion length in units of bead diameters, $b^{-1}\sqrt{D_0 t}$, where $b$ is the bead diameter. The calculated limits at short-$t$ ($S/V_p$) and long-$t$ (tortuosity) are shown by the dashed and solid lines respectively, for an idealized sample of porosity $\phi = 0.38$, $1/\alpha = \sqrt{\phi} = 0.62$, $S/V_p = 6(1-\phi)/(\phi b) = 9.79/b$. (a) Xenon diffusion in packs of 1, 2, 3 and 4 mm beads. (b) Xenon diffusion in packs of 0.1, 0.5 and 1 mm beads, showing diffusion over distances exceeding many bead diameters. Example error bars for all samples are shown only for the 0.1 mm beads data series.

Fig. 4. $^{129}$Xe time-dependent diffusion measurements, $D(t)$, for two different xenon gas pressures in samples of randomly packed spherical glass beads. $D(t)$ was measured at imbibed Xe gas pressures of ~ 6 bar (black squares) and 3 bar (white squares). The 3 bar pressure data is reproduced from ref. (*30*). For 0.1, 0.5, 1 and 4 mm bead samples, water $D(t)$ data is also shown (white circles). The water data for 0.1 mm beads is reproduced from ref. (*1*), but otherwise was measured for this work. As in Fig. 3,

$D(t)/D_0$ is plotted against $b^{-1}\sqrt{D_0 t}$, in bead diameters. The short-$t$ and long-$t$ limits for $D(t)/D_0$ are derived from the 6 bar Xe data, and are shown by the dashed and solid lines respectively. For each sample, the tortuosity is derived from the measured $D(t)/D_0$ asymptote, and then related to the porosity ($1/\alpha \sim \sqrt{\phi}$), from which the expected $S/V_p$ is derived in order to plot the dashed line. Additionally for each plot, the Pade′ approximation, shown by a curved line, has been evaluated (Eq. [4]) using the value of $\sqrt{D_0 \theta}$ that extrapolates to the long and short-$t$ limits over the longest diffusion distance. Error bars are shown on the ~ 6 bar data only when they significantly exceed the size of the data symbols. (a) 4 mm bead sample. (b) 3 mm bead sample. (c) 2 mm bead sample. (d) 1 mm bead sample. (e) 0.5 mm bead sample. (f) 0.1 mm bead sample.

Fig. 5. Exploration of finite sample size effects in $^{129}$Xe time-dependent diffusion measurements, $D(t)$. Randomly packed spherical 4 mm diameter glass beads were held in two different sample containers: a small cell of ~ 50 cm$^3$ (black squares) and a large cell of ~ 150 cm$^3$ (white squares). For both samples, $D(t)$ was measured at imbibed Xe gas pressures of ~ 6.5 bar. Once again, $D(t)/D_0$ is plotted against $b^{-1}\sqrt{D_0 t}$, in bead diameters. The data from the small cell, the short-$t$ and long-$t$ limits for $D(t)/D_0$, as well as the Pade′ approximation, are the same as that shown in Fig. 4a) for 6.41 bar of Xe. In the larger sample cell the Xe $D(t)/D_0$ data agrees well with the Pade′ approximation approach to the tortuosity limit. However, in the smaller cell, a significant deviation from the Pade′ line can be seen: the tortuosity limit is reached earlier. Thus finite sample size effects can induce a systematic underestimate of $L_{macro}$, the largest structural length scale in the pore geometry.

Fig. 6. Hypothesized three phenomena that can effect the measurement of $D(t)/D_0$ using the PGStE-bp NMR technique. The effect of spin movement during $\delta$ (phenomenon 2) in the measurement of free diffusion, and in the case or restricted diffusion (phenomena 1 and 2 combined) is well known. Our current study presents a sample in which all three phenomena occur. $l_b$ refers to the characteristic length scale of the bead, which could be taken as either the bead diameter or pore size between beads.

| bead diameter(mm) | $\phi$ | $1/\alpha$ | $b\ (S/V_p)$ | $\theta$ (s) | $D_0$ (m$^2$s$^{-1}$) | $b^{-1}\sqrt{D_0\theta}$ |
|---|---|---|---|---|---|---|
| *Xenon gas* | | | | | | |
| $b = 4$ | 0.38 | 0.62 | 9.79 | 0.34 | 8.08×10$^{-7}$ | 0.131 |
| $b = 3$ | 0.38 | 0.62 | 9.79 | 0.18 | 8.12×10$^{-7}$ | 0.127 |
| $b = 2$ | 0.39 | 0.63 | 9.19 | 0.085 | 8.20×10$^{-7}$ | 0.132 |
| $b = 1$ | 0.41 | 0.64 | 8.42 | 0.022 | 8.13×10$^{-7}$ | 0.134 |
| $b = 0.5$ | 0.42 | 0.65 | 8.22 | 0.006 | 7.92×10$^{-7}$ | 0.138 |
| $b = 0.1$ | 0.44 | 0.66 | 7.57 | 0.00022 | 7.88×10$^{-7}$ | 0.132 |
| *Water* | | | | | | |
| $b = 0.1$ | 0.41 | 0.64 | 8.65 | 0.12 | 1.76×10$^{-9}$ | 0.145 |
| $b = 0.05$ | 0.41 | 0.64 | 8.65 | 0.03 | 1.77×10$^{-9}$ | 0.146 |

**Table 1**

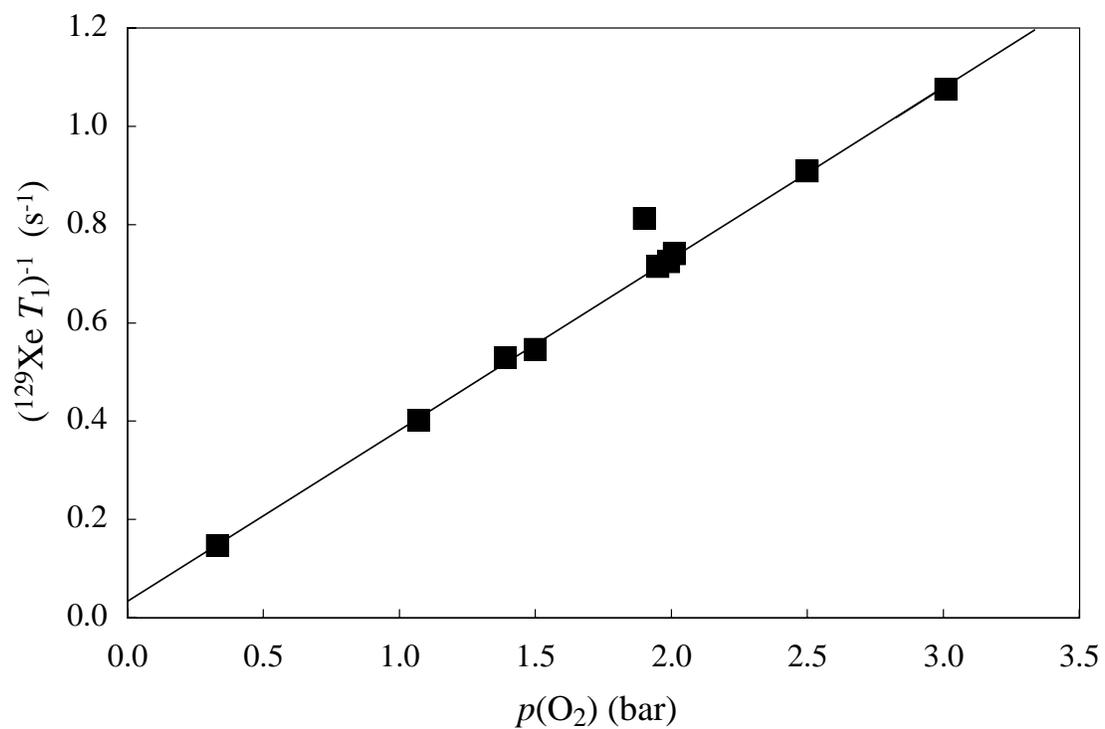

**Figure 1**

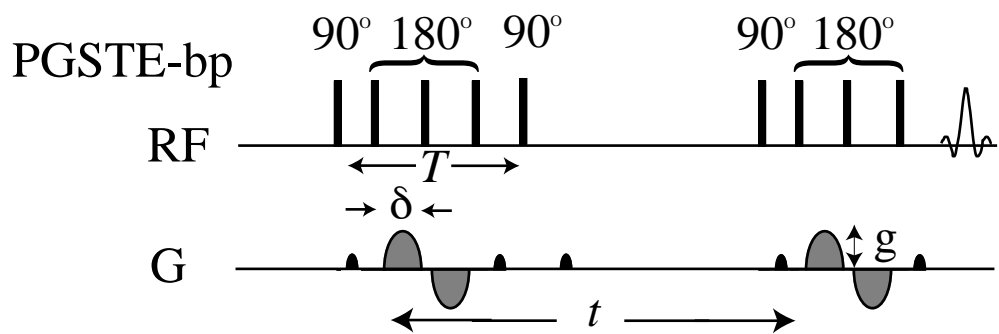

**Figure 2**

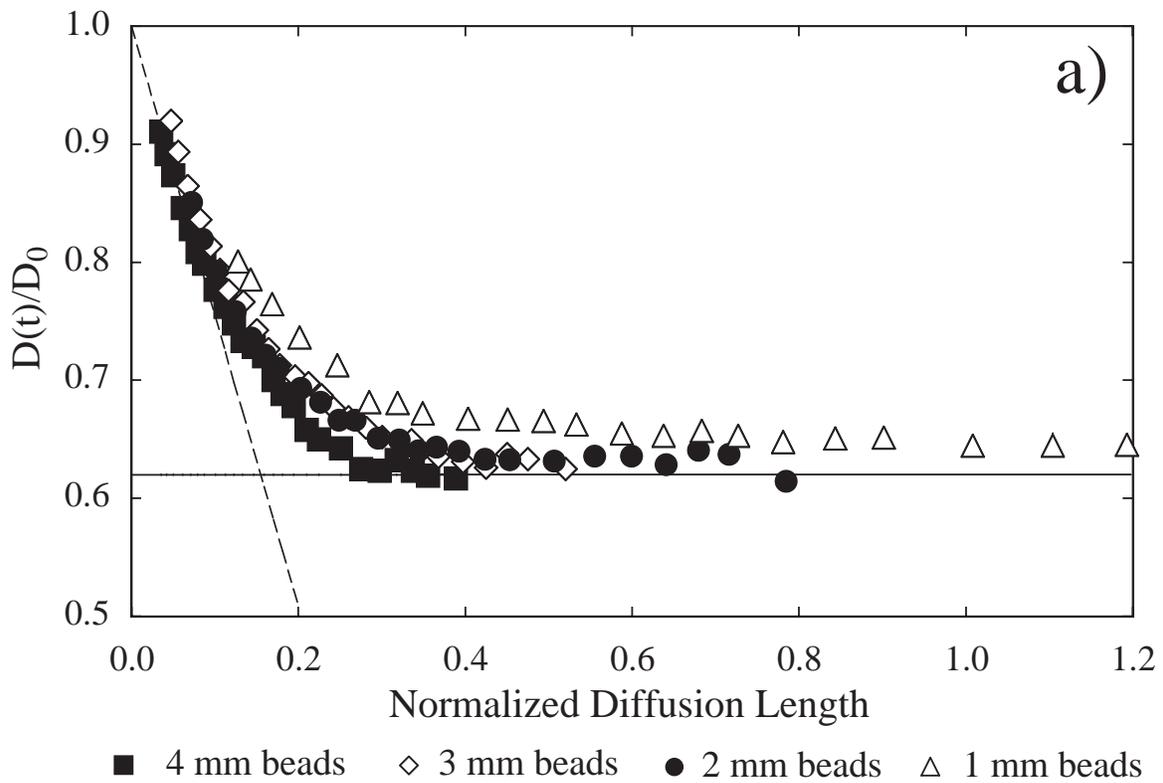
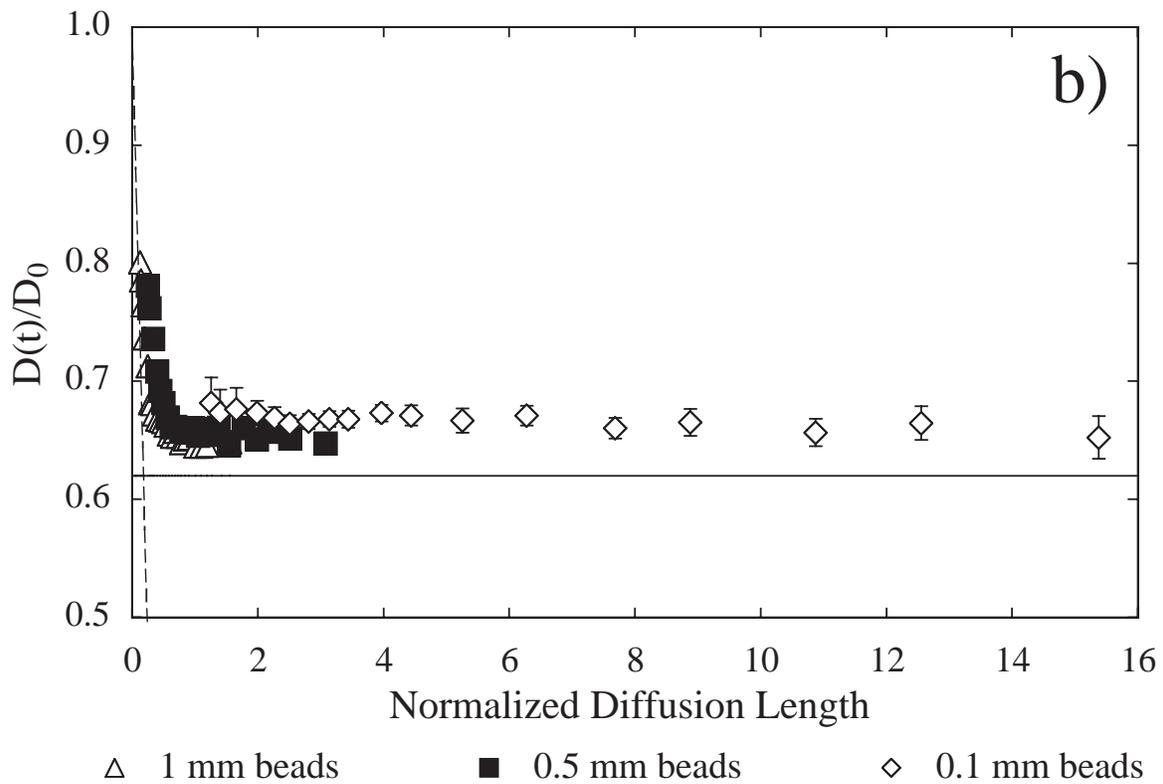

**Figure 3**

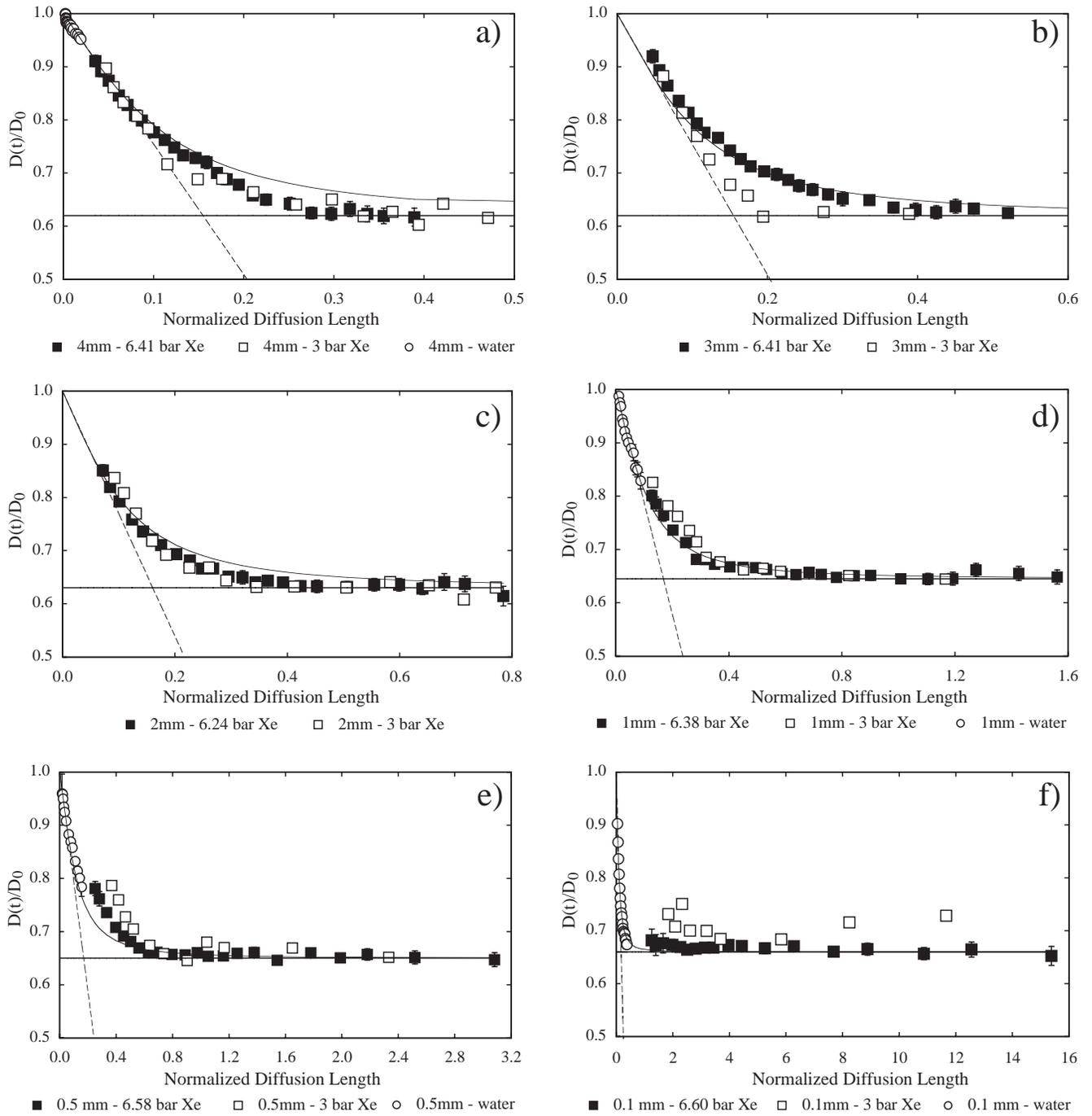

**Figure 4**

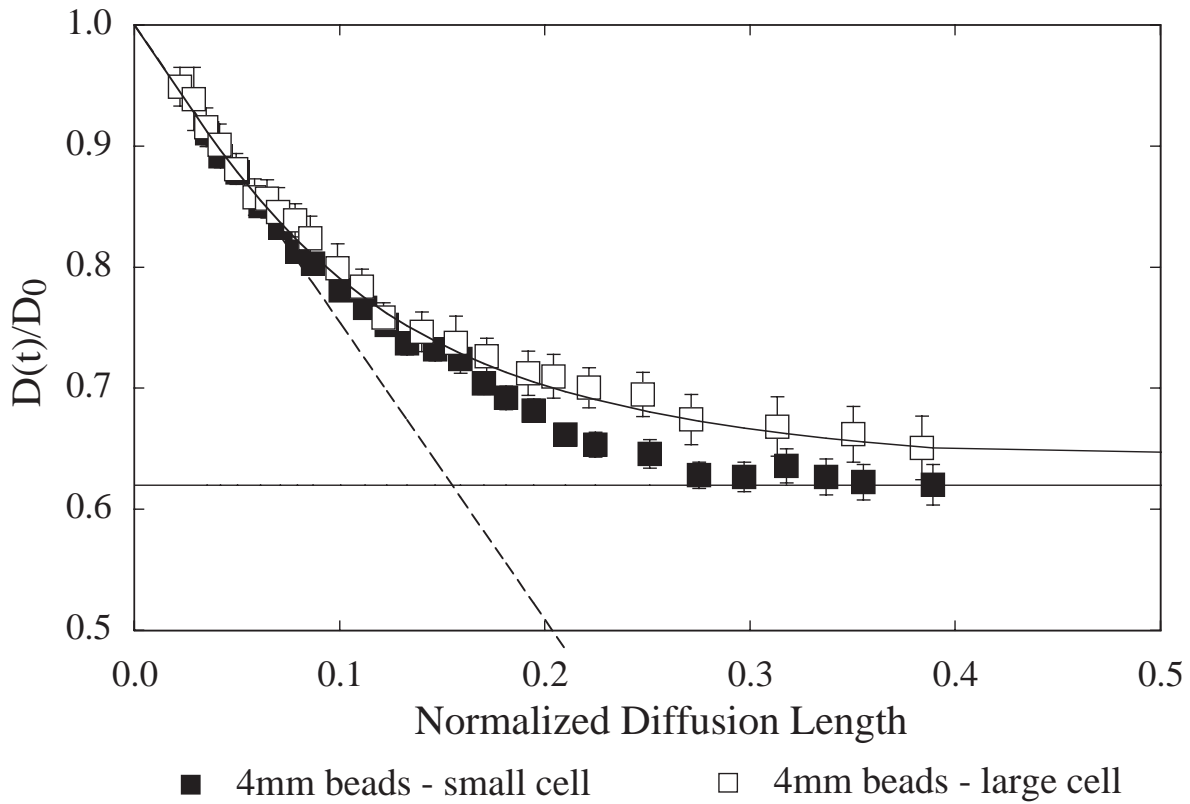

**Figure 5**

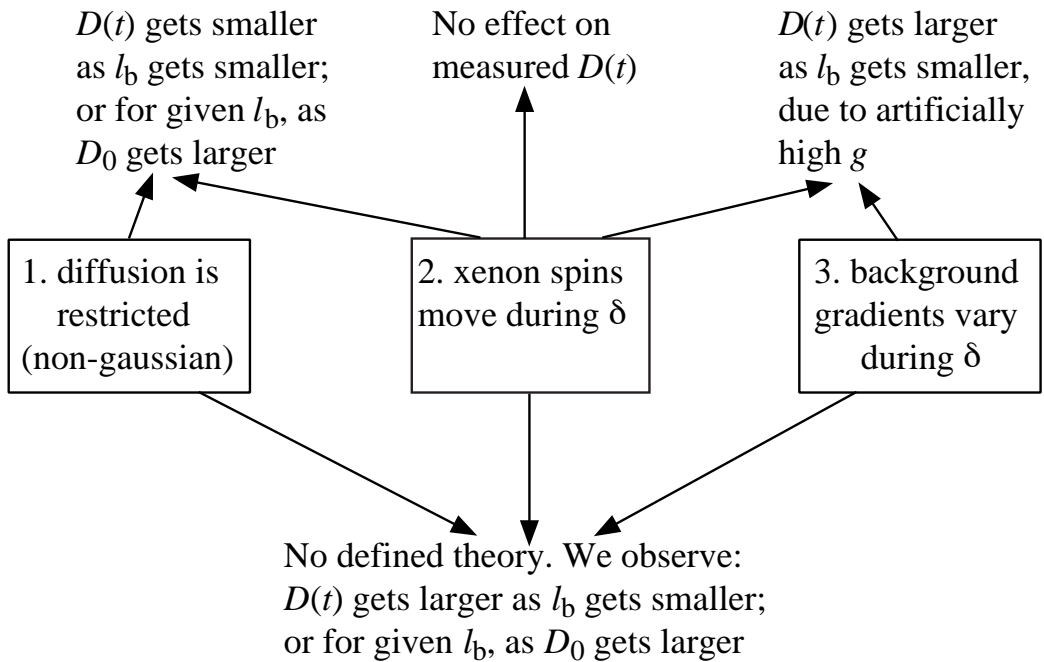

**Figure 6**